\documentclass[prd,twocolumn,superscriptaddress,nofootinbib]{revtex4-1}

\usepackage{amsmath,amssymb}
\usepackage{epsfig}
\usepackage{graphicx}
\usepackage{mathrsfs}
\usepackage{amsmath}
\usepackage{amsfonts}
\usepackage{epstopdf}
\usepackage{color}

\usepackage{tikz}
\usepackage{pgfplots}
\usepgfplotslibrary{external}
\tikzset{external/system call={lualatex --shell-escape
		\tikzexternalcheckshellescape --halt-on-error --interaction=batchmode
		--jobname "\image" "\texsource"}}
\tikzset{cross/.style={cross out, draw=black, minimum size=2*(#1-\pgflinewidth), inner sep=0pt, outer sep=0pt},
	cross/.default={2.2pt}}
\usetikzlibrary{pgfplots.groupplots}

\renewcommand{\Im}{\mathfrak{Im}}
\newcommand{\dd}{\mathrm{d}}
\newcommand{\psibar}{\bar{\psi}}

\newcommand{\Tr}{\mathrm{Tr}}
\newcommand{\dr}{\partial}

\newcommand{\ma}[2]{a_{#1,#2}}
\newcommand{\madag}[2]{a_{#1,#2}^\dagger}
\newcommand{\mb}[2]{b_{#1,#2}}
\newcommand{\mbdag}[2]{b_{#1,#2}^\dagger}

\newcommand{\malpha}[2]{\alpha_{#1,#2}}
\newcommand{\mbeta}[2]{\beta_{#1,#2}}
\newcommand{\mgamma}[2]{\gamma_{#1,#2}}
\newcommand{\mdelta}[2]{\delta_{#1,#2}}

\newcommand{\spinorp}[2]{\chi^+_{#1,#2}}
\newcommand{\spinorm}[2]{\chi^-_{#1,#2}}
\newcommand{\spinorpm}[2]{\chi^\pm_{#1,#2}}

\newcommand{\spinorpbar}[2]{\bar{\chi}^{+}_{#1,#2}}
\newcommand{\spinormbar}[2]{\bar{\chi}^{-}_{#1,#2}}

\newcommand{\vacstate}[1]{\left|\Omega_{#1}\right\rangle}
\newcommand{\vacstatebra}[1]{\left\langle\Omega_{#1}\right |}

\newcommand{\twopart}[4]{\left|#1^a_{#3,#4},#2^b_{-#3,#4}\right\rangle}
\newcommand{\twopartbra}[4]{\left\langle#1^a_{#3,#4},#2^b_{-#3,#4}\right|}
\newcommand{\twopartbracket}[4]{\left\langle#1^a_{#3,#4},#2^b_{-#3,#4}\right|\left.#1^a_{#3,#4},#2^b_{-#3,#4}\right \rangle}

\newcommand{\rrho}[6]{\twopart{#1}{#2}{#5}{#6}\twopartbra{#3}{#4}{#5}{#6}}

\newcommand{\toto}{\frac{t}{\tau}}
\newcommand{\totop}{\left(\toto\right)}
\newcommand{\Ec}{\mathcal{E}}

\usepackage{tensor}
\newcommand{\hyp}[4]{\ \tensor[_2]{F}{_1}\left(#1,#2,#3,#4\right)}

\begin{document}

\title{Gibbs entropy from entanglement  in electric quenches}

\author{Adrien Florio}
\email{adrien.florio@stonybrook.edu}
\affiliation{Center for Nuclear Theory, Department of Physics and Astronomy, Stony Brook University, New York 11794-3800, USA}
\author{Dmitri E. Kharzeev}
\email{Dmitri.Kharzeev@stonybrook.edu}
\affiliation{Center for Nuclear Theory, Department of Physics and Astronomy, Stony Brook University, New York 11794-3800, USA}
\affiliation{Department of Physics,
Brookhaven National Laboratory, \\ Upton, New York 11973-5000, USA}

\date{\today}


\begin{abstract}
In quantum electrodynamics with charged  fermions, a background electric field is the source ofthe chiral anomaly which creates a chirally imbalanced state of fermions. This chiral state is realized through the production of entangled pairs of  right-moving fermions and left-moving antifermions (or vice versa, depending on the orientation of the electric field).
Here we show that the statistical Gibbs entropy associated with these pairs is equal to the entropy of entanglement between the right-moving particles and left-moving antiparticles.  We then derive an asymptotic expansion for the entanglement entropy in terms of the cumulants of the multiplicity distribution of produced particles and explain how to re-sum this asymptotic expansion. Finally, we study the time dependence of the entanglement entropy in a specific time-dependent pulsed background electric field, the so-called "Sauter pulse", and illustrate how our re-summation method works in this specific case. We also find that short pulses (such as the ones created by high energy collisions) result in an approximately thermal distribution for the produced particles.
\end{abstract}
\maketitle

\section{Introduction}

The notion of entanglement played a key role in the development \cite{Einstein:1935rr} and validation \cite{Aspect:1982fx} of quantum mechanics.
It also plays a crucial role in the rapidly developing field of quantum computing, where entangled qubits may provide an exponential improvement over classical computers, see e.g. \cite{Nielsen:2011}.

The precise role played by entanglement in quantum field theory is, however, still an open question.  One of the first contributions in this direction was made in \cite{Srednicki:1993im}, where it was shown that the entanglement entropy between a massless free field inside an imaginary sphere and the rest of the system (described as a "bath"), reproduces the famous "area law" of black hole thermodynamics. In a parallel development, the spatial dependence of the entanglement entropy in conformal field theories was established, with its logarithmic dependence on the size of the subregion \cite{Holzhey:1994we,hep-th/0405152}, see \cite{Calabrese:2009qy} for a review. The link between  entanglement, quantum chaos, and thermalization is under active current investigation, see for example \cite{Maldacena:2013xja,Asplund:2015eha, Cotler:2016acd, Calabrese:2016xau,PhysRevD.100.026002, Mezei:2018jco}  and references therein.

Another question that has recently sparked interest is the extent to which the renormalization group flow, which connects high (UV) and low (IR) energies, can be understood in terms of momentum-space entanglement between the UV and the IR degrees of freedom \cite{Balasubramanian:2011wt,Han:2020uwn,Klco:2021biu}. It also starts to become clear that the concept of entanglement in quantum field theory is not only of academic interest but can have important implications, see e.g.  \cite{Kharzeev:2005iz,lin2009grazing,bacsar2012holographic,Ho:2015rga,Muller:2017vnp,Kharzeev:2017qzs,Berges:2017hne,Baker:2017wtt,Tu:2019ouv,Kharzeev:2021yyf} where diverse ideas related to entanglement were put forward in an attempt to understand thermalization in high energy heavy-ion collisions.

High energy hadron collisions are accompanied by the production of copious quark-antiquark pairs. To investigate the role that quantum entanglement plays in this process, here
we will study the entanglement entropy between  fermions and antifermions produced in strong electric fields.
For simplicity, we deal with  $1+1$ dimensional electrodynamics; however, our results can be easily extended to $3+1$ dimensions as well.

The main two results of this paper are the following.  First, we show that the statistical Gibbs entropy associated with the created pairs is equal to the entanglement entropy between right and left movers (we refer to this as "chiral entanglement"). This result elucidates the microscopic quantum origin of the statistical entropy of the produced state. Second, we derive an explicit relation between the entanglement entropy and the multiplicity distribution of created particles and show how it can be used in practice to reconstruct the entanglement entropy from the knowledge of the (few) cumulants of the multiplicity distribution.

To this aim, we start in section \ref{sec:gibbsEntropy} by rederiving the results of \cite{Ebadi:2014ufa} and compute the entanglement entropy between particles in a pair created by a background electric field. We then compare it to the  Gibbs entropy associated with the system of the created pairs and show them equal. We then move on to section \ref{sec:cummulants} and relate the entanglement entropy to the cumulants of the multiplicity distribution of produced particles. The asymptotic expansion we find is the same as the one previously derived in the context of shot noise in quantum point contacts (``full counting statistics") \cite{Klich:2008un}. We also show how to re-sum this expansion for practical applications. Then, in section~\ref{sec:sauterField}, we study the time evolution of the entanglement entropy for a particular time dependence of the electric pulse ("Sauter pulse"). We observe that in the case of short pulses,  the asymptotic values are approached in a universal way. Finally, we show how our re-summed expression can be used in practice to evaluate the entanglement entropy from the multiplicity distribution of produced particles. In particular, we show that it provides a substantial improvement over the original asymptotic expansion. We conclude in section \ref{sec:conclu} by summarizing our results and presenting an  outlook.

\section{Chiral entanglement and Gibbs entropy}
\label{sec:gibbsEntropy}

For the clarity of the argument, we will consider $1+1D$ massive fermions coupled to a background electric field
\begin{align}
  S= \int \dd^2 x\ \hat\psibar \left ( i\gamma^\mu\partial_\mu +A_1(t)\gamma^1 - m\right)\hat\psi \ . \label{eq:DiracAc}
\end{align}
with $\gamma^\mu$ the Dirac matrices, $A_1(t)$ a homogeneous, time-dependent background gauge field, working in the $(+,-)$ signature and with $\psibar = \psi^\dagger \gamma^0$ the usual Dirac conjugate.

We emphasize that this choice is out of convenience;  the computations can also be carried for free fermions in a purely electric background in $3+1$ dimensions. In this case, one needs to introduce a transverse momentum $p_T$ for each pair, which then needs to be integrated over; we refer the interested  reader to \cite{Dunne:2004nc, Ruffini:2009hg}
for more information on pair creation in $3+1$ dimensions.  Throughout this work, we will be concerned with pair creation, associated entropies, and thermalization. To tackle these questions, we will use the method of Bogoliubov transformations \cite{bogoliubov1947theory}, see \cite{Birrell:1982ix,Ambjorn:1983hp,Blaschke:2017igl,Domcke:2021fee} for examples and reviews. It will be particularly useful here because it will allow us to explicitly construct and work with a Fock space, giving us a direct way to compute density matrices and entropies thereof.

At any instant of time $t^*$, the Hamiltonian associated to \eqref{eq:DiracAc}
\begin{align}
  H_{t^*}=\int\dd x_1 \hat\psibar(t^*,x_1)\left(-\gamma^1\left(i\dr_1+A_1(t^*)\right)+m\right)\hat\psi(t^*,x_1)
\end{align}
is diagonalized by introducing creation and annihilation operators $\ma{k_1}{t^*}, \madag{k_1}{t^*}, \mb{k_1}{t^*}, \mbdag{k}{t^*}$ satisfying the usual anticommutation relations
\begin{align}
  \{\madag{k_1}{t^*},\ma{q_1}{t^*}\}&=\{\mbdag{k_1}{t^*},\mb{q_1}{t^*}\} = 2\pi \delta(k_1-q_1) \\
\{\ma{k_1}{t^*},\mb{q_1}{t^*}\}&=   \{\madag{k_1}{t^*},\mb{q_1}{t^*}\} = 0 \ .
\label{eq:anticommutation}
\end{align}
 Indeed, expanding the field operator as
\begin{align}
  \hspace{-0.1cm}\hat\psi(t,x_1)&=\int \dd k_1 e^{-ik_1 x^1}\left(\spinorp{k_1}{t^*}(t)\ma{k_1}{t^*}+\spinorm{-k_1}{t^*}(t)\mbdag{k_1}{t^*}\right) \label{eq:modesExpansion}\\
    \hspace{-0.1cm}\hat\psibar(t,x_1)&=\int \dd k_1 e^{ik_1 x^1}\left(\spinorpbar{k_1}{t^*}(t)\madag{k_1}{t^*}+\spinormbar{-k_1}{t^*}(t)\mb{k_1}{t^*}\right) \ ,
\end{align}
with $\spinorpm{\pm k_1}{t^*}(t)$ solutions to the Dirac equation for the momenta modes
\begin{align}
  \left(i\gamma^0\dr_0  + \gamma^1(k_1+A_1(t))-m\right)\spinorpm{\pm k_1}{t^*}(t) = 0 \ , \label{eq:DiracEq}
\end{align}
it is easy to see that $H_{t^*}$ is diagonal, provided that the mode functions at time $t^*$, $\spinorpm{\pm k_1}{t^*}(t^*) $ reduce to the free solutions $u^\pm_{A_1}$  of the Dirac equation \eqref{eq:DiracEq} in a \textit{time-independent} background $A_1$;
\begin{align}
\spinorpm{\pm k_1}{t^*}(t^*) = u^\pm_{A_1(t^*)}(t^*,\pm k_1) \ .
\end{align}
For an explicit form of the free  spinors $u^\pm_{A_1}$, see appendix \ref{app:freeSpinors}.

The fact that these creation and annihilation operators diagonalize the Hamiltonian at time $t^*$ means that they define the concept of free particles at $t=t^*$. Conversely, they also define the notion of  the vacuum state at $t=t^*$; it is the state annihilated $\vacstate{t^*}$ by $\ma{k_1}{t^*}$ and $\mb{k_1}{t^*}$
\begin{align}
\ma{k_1}{t^*}\vacstate{t^*}=\mb{k_1}{t^*}\vacstate{t^*}=0 \ .
\end{align}
In this language, particle creation directly comes from the fact that the vacuum is not uniquely defined; what is the vacuum state at some time will correspond to some excited state at some later time. We will consider situations with $\lim_{t\to-\infty}A_1(t)=A_{-\infty}$, namely backgrounds where the electric field is switched-off at $t=-\infty$, and ask what is the particle content of this initial vacuum state $\vacstate{-\infty}$ at time $t=t^*$. To do so, we relate the asymptotic operators to the ones at time $t^*$ using a Bogoliubov transformation
\begin{align}
  \ma{k_1}{t^*}=\malpha{k_1}{t^*}\ma{k_1}{-\infty}+\mbeta{k_1}{t^*}^*\mbdag{-k_1}{-\infty} \label{eq:bogoa}\\
  \mb{-k_1}{t^*}=\mdelta{k_1}{t^*}\mb{-k_1}{-\infty}+\mgamma{k_1}{t^*}^*\madag{k_1}{-\infty} \ . \label{eq:bogob}
\end{align}
As the anticommutation relations \eqref{eq:anticommutation} must be satisfied at all times, we have the following constraints
\begin{align}
|\malpha{k_1}{t^*}|^2+|\mbeta{k_1}{t^*}|^2=1, \ \ |\mdelta{k_1}{t^*}|^2&+|\mgamma{k_1}{t^*}|^2=1, \\
\malpha{k_1}{t^*}\mgamma{k_1}{t^*}^*+\mbeta{k_1}{t^*}^*\mdelta{k_1}{t^*}&=0
\end{align}
and find that only two out of the four coefficients are independent. Consistently with the constraints and the fact that $\malpha{k_1}{-\infty}=\mdelta{k_1}{-\infty}=1, \mbeta{k_1}{-\infty}=\mgamma{k_1}{-\infty}=0$ , we set
\begin{align}
  \mdelta{k_1}{t^*} &=   \malpha{k_1}{t^*} \ , \ \ \mgamma{k_1}{t^*} = -\mbeta{k_1}{t^*} \ . \label{eq:constraintsBogo}
\end{align}
These coefficients can readily be extracted from the knowledge of the mode functions $\spinorpm{k_1}{-\infty}$. First we plug the transformations~\eqref{eq:bogoa}-\eqref{eq:bogob}, taking into account the constraints \eqref{eq:constraintsBogo}, into \eqref{eq:modesExpansion}. Regrouping by operator and comparing with the field expansion \eqref{eq:modesExpansion} around $t^*=-\infty$, we find
\begin{align}
\spinorp{k_1}{t^*}(t^*) = u^+_{A_{-\infty}}(t^*,k_1)\malpha{k_1}{t^*} - u^-_{A_{-\infty}}(t^*,-k_1)^\dagger\mbeta{k_1}{t^*} \label{eq:bogoModes}\ .
\end{align}
We can then use the normalization \eqref{eq:freeNorm} of the free spinors together with their orthonormality relations \eqref{eq:freeOrtho} to invert this relation. This allows us the compute the Bogoliubov coefficients as
\begin{align}
\malpha{k_1}{t^*}&=u^{+\dagger}_{A_{-\infty}}(t^*,k_1)\spinorp{k_1}{t^*}(t^*) \label{eq:alphaFromModes}\\
\mbeta{k_1}{t^*}&=- u^{-\dagger}_{A_{-\infty}}(t^*,-k_1)\spinorp{k_1}{t^*}(t^*) \ .\label{eq:betaFromModes}
\end{align}
Note that one can also get the Bogoliubov coefficients by considering \eqref{eq:bogoModes} and its derivative with respect to time and taking linear combination \cite{Blaschke:2017igl}. It is easy to show that both methods are equivalent on-shell; the one presented here is more straightforward once the mode functions are known.

The meaning of the Bogoliubov coefficients is clear. In particular, we have
\begin{align}
\vacstatebra{-\infty }\madag{k_1}{t^*}\ma{k_1}{t^*}\vacstate{-\infty}&=\vacstatebra{-\infty }\mbdag{-k_1}{t^*}\mb{-k_1}{t^*} \vacstate{-\infty}\notag\\
&=|\mbeta{k_1}{t^*}|^2
\end{align}
and  $|\mbeta{k_1}{t^*}|^2$ corresponds to the probability of observing at time $t^*$, having started out in the vacuum state, a(n) (anti)particle with momentum $(-)k$.

This interpretation allows us to associate a statistical Gibbs entropy to the system at time $t^*$ by counting  microstates. As we are dealing with fermions, only one particle per momentum can be excited. Thus
\begin{align}
  S_G = \int \frac{\dd k_1}{2\pi} \big[&\left(1-|\mbeta{k_1}{t^*}|^2\right)\log\left( 1-|\mbeta{k_1}{t^*}|^2\right) \notag\\
  &+|\mbeta{k_1}{t^*}|^2\log\left(|\mbeta{k_1}{t^*}|^2\right)\big] \label{eq:GibbsEntropy} \ ,
\end{align}
as every momentum is either excited or not.

We will now show that the entanglement entropy between positive and negative frequency modes, or equivalently between left- and right-moving (anti)fermions, as computed from first principles, is equal to the Gibbs entropy \eqref{eq:GibbsEntropy}, elucidating the microscopic origin of this quantity.

The computation of the entanglement entropy was already essentially presented in \cite{Ebadi:2014ufa}. For the sake of clarity and self-consistency, we also present it here. In terms of one-particle states, the vacuum reads
\begin{align}
  \vacstate{-\infty}= \left|L\right\rangle \otimes \left|R\right\rangle\label{eq:rawvac}
\end{align}
where
\begin{align}
  \left|L\right\rangle &=
  \bigotimes_{k_1}\left|0^a_{k_1}\right\rangle\\
  \left|R\right\rangle &=\bigotimes_{q_1}\left|0^b_{q_1}\right\rangle
\end{align}
and with the index $a$ resp. $b$ referring to the particle resp. antiparticle Hilbert's space. We used the notation $\otimes$ to indicate that we are dealing with tensor products. Conservation of charge, translational invariance,  and time-reversal invariance  lead us to express the vacuum state as a tensor products of pairs $\twopart{0}{0}{k_1}{-\infty}$ with a spectator state $|Sp\rangle$
\begin{align}
  \vacstate{-\infty}=\left(\bigotimes_{k_1} \twopart{0}{0}{k_1}{-\infty}\right)\otimes|Sp\rangle\label{eq:lessrawvac} \ .
\end{align}
The state $|Sp\rangle$ contains all the combinations which cannot be written as pairs and is defined such that expressions \eqref{eq:rawvac} and \eqref{eq:lessrawvac} are the same. As turning on a homogeneous background electric field preserves all these symmetries, this remainder term will not evolve and will not impact the discussion.

 We start from the density matrix associated with the vacuum state
\begin{align}
  \rho|_{t^*}=\vacstate{-\infty}\vacstatebra{-\infty}=\left(\bigotimes_{k_1} \rho_{k_1,-k_1}|_{t^*}\right)\otimes |Sp\rangle\langle Sp|
\end{align}
with
\begin{align}
  \rho_{k_1,-k_1}|_{t^*}=\rrho{0}{0}{0}{0}{k_1}{-\infty} \ .
\end{align}
The subscript $|_{t^*}$ is here to remind us we want to evaluate these quantities at time $t^*$, namely in the "instantaneous basis" $\ma{k_1}{t^*}, \mb{k_1}{t^*}$. We also define a reduced density matrix by tracing over half of the particles and the spectator state
\begin{align}
  \rho^+|_{t^*}=\Tr_{-k_1,Sp}\left(\rho|_{t^*}\right) =  \bigotimes_{k_1} \rho_{k_1}|_{t^*}
\end{align}
with
\begin{align}
  \rho_{k_1}|_{t^*}=\Tr_{-k_1}\left(\rho_{k_1,-k_1}|_{t^*}\right) \ .
\end{align}
We want to compute the entanglement entropy associated to this reduced density matrix
\begin{align}
  S_E = -\Tr\left(\rho^+|_{t^*}\log\left(\rho^+|_{t^*}\right)\right) =\int\frac{\dd k_1}{2\pi} S_k \ ,
\end{align}
with
\begin{align}
  S_k = \Tr\left(\rho_{k_1}|_{t^*}\log\left(\rho_{k_1}|_{t^*}\right)\right)\ .
\end{align}

Our aim now is to compute $S_k$. To do so, we expand our state in the instantaneous basis. Conservation of charge leads us to write
\begin{align}
  \twopart{0}{0}{k_1}{-\infty} = \lambda_0 &\twopart{0}{0}{k_1}{t^*} \notag \\
  &\phantom{aa}+  \lambda_1 \twopart{1}{1}{k_1}{t^*} \ , \label{eq:vactransfo}
\end{align}
leading to
\begin{widetext}
  \begin{align}
\rho_{k,-k}&=  |\lambda_0|^2\rrho{0}{0}{0}{0}{k_1}{t^*} +  |\lambda_1|^2 \rrho{1}{1}{1}{1}{k_1}{t^*} \\
&\phantom{aaaaaaaaaaaaaaaaaaaaa}+\lambda_0\lambda_1^* \rrho{0}{0}{1}{1}{k_1}{t^*} +\lambda_0^*\lambda_1 \rrho{1}{1}{0}{0}{k_1}{t^*}\notag \\
\rho_{k}&=  |\lambda_0|^2\left|0^a_{k_1,t^*}\right\rangle\left\langle0^a_{k_1,t^*}\right| +  |\lambda_1|^2 \left|1^a_{k_1,t^*}\right\rangle\left\langle1^a_{k_1,t^*}\right|
\end{align}

\end{widetext}
and
\begin{align}
  S_k&=-\left(|\lambda_0|^2\log\left(|\lambda_0|^2\right))+|\lambda_1|^2\log\left(|\lambda_1|^2\right))\right) \ . \label{eq:Sk}
\end{align}
The only missing piece is to relate the $\lambda$'s to the Bogoliubov coefficients. This is done by inverting the transformation \eqref{eq:bogoa}-\eqref{eq:bogob}
\begin{align}
  \ma{k_1}{-\infty}&=\malpha{k_1}{t^*}^*\ma{k_1}{t^*}-\mbeta{k_1}{t^*}^*\mbdag{-k_1}{t^*} \label{eq:bogoinva}\\
  \mb{-k_1}{-\infty}&=\malpha{k_1}{t^*}^*\mb{-k_1}{t^*}+\mbeta{k_1}{t^*}^*\madag{k_1}{t^*} \ . \label{eq:bogoinvb}
\end{align}
By definition of the vacuum state, we need to have
\begin{align}
  \ma{k_1}{-\infty} \twopart{0}{0}{k_1}{-\infty}\stackrel{!}{=}0
\end{align}
which implies that, after plugging in equations \eqref{eq:vactransfo} and \eqref{eq:bogoinva},
\begin{align}
  \malpha{k_1}{t^*}^*\lambda_1 = \mbeta{k_1}{t^*}^*\lambda_0 \ . \label{eq:lambacond1}
\end{align}
 One can check one finds the same condition by imposing $\mb{k_1}{-\infty} \twopart{0}{0}{k_1}{-\infty}\stackrel{!}{=}0$. We can obtain a second consistency condition using the normalisation of the vacuum state
\begin{align}
   \twopartbracket{0}{0}{k_1}{-\infty}\stackrel{!}{=}1 \ .
\end{align}
Using again the definitions \eqref{eq:vactransfo} and \eqref{eq:bogoinva} we find
\begin{align}
  |\lambda_0|^2+|\lambda_1|^2=1
\end{align}
and together with \eqref{eq:lambacond1} it implies
\begin{align}
    |\lambda_0|^2 = \left|\malpha{k_1}{t^*}\right |^2 \ \ , \ \   |\lambda_1|^2=\left|\mbeta{k_1}{t^*}\right |^2 \ .
\end{align}
Using equation \eqref{eq:Sk}, we obtain
\begin{align}\label{eq:EEntropy}
  S_E=-\int \frac{\dd k_1}{2\pi}\Big[\left|\malpha{k_1}{t^*}\right |^2& \log\left(\left|\malpha{k_1}{t^*}\right |^2 \right)\\
  &+\left|\mbeta{k_1}{t^*}\right |^2\log\left(\left|\mbeta{k_1}{t^*}\right |^2\right)\Big]\notag \ .
\end{align}
Note that this expression was already derived in  \cite{Ebadi:2014ufa}.

Comparing this expression with (\ref{eq:GibbsEntropy}), and
using $\left|\malpha{k_1}{t^*}\right |^2=1-\left|\mbeta{k_1}{t^*}\right |^2$, we find
\begin{align}
S_G = S_E \ .
\end{align}
We stress again that this result provides a microscopic interpretation of the thermodynamic entropy associated with the particles created: it arises from the microscopic quantum theory as the entropy of entanglement between the particles in a produced pair. The apparent statistical behavior is a reflection of the quantum entanglement. Note also that this relation between the Gibbs entropy and the entanglement entropy also holds in the case of a scalar particle, as can be seen from the expression derived for the latter in \cite{Ebadi:2014ufa}.

\phantom{space}

\section{Relation between the entanglement entropy and the multiplicity distribution}
\label{sec:cummulants}

The fact that the particles emitted in pairs are entangled must be directly reflected in the multiplicity distribution of created particles. Here, we will make this link explicit and directly express the entanglement entropy in terms of the generating function for the multiplicity distribution of particles. This relation seems to be rather general as it was already derived in the context of quantum shot noise \cite{Klich:2008un}; here we derive it for arbitrary time-dependent pulses.

Let us first regularize our theory by temporarily putting it on a circle of length $R$ (we will then take the limit $R\to\infty$). As a result, the momenta are now quantized $p_l=\frac{2\pi l}{R}$. Let us define $P_n$ to be the probability of creating $n$ fermions. Due to Fermi's exclusion principle, it can be written as
\begin{align}
  P_n =\sum_{\substack{\alpha \\ |I_\alpha|=n}} \prod_{l\in I_\alpha} n_{p_l} \prod_{q\notin I_\alpha} (1-n_{p_q})
  \label{eq:probNpairs}
\end{align}
with $n_{p_l}$ the probability to create a particle of momentum $p_l$ and $I_\alpha$ a set of distinct indices. We use the notation $|I_\alpha|$ to denote the cardinal of $I_\alpha$, i.e. the number of indices. This formula is understood as follows: for any combination of $n$ momenta, we require these $n$ momenta to contain a particle (first product) and all other momenta to be empty (second product).

A more concise way to encode this information is obtained by considering its associated generating function
\begin{align}
  M(\lambda)=\sum_{n=0}^\infty P_n e^{i\lambda n}\ . \label{eq:chidef}
\end{align}
It is a generating function in the sense that the $m^{th}$ moment of the distribution of the number of created particles can be obtained by taking derivatives of $\log(M(\lambda))$.

By using the explicit definition for the probability of creating $n$ particles \eqref{eq:probNpairs}, we can rewrite this quantity as follows
\begin{align}
  M(\lambda)&=\sum_{n=0}^\infty  e^{i\lambda n}\sum_{\substack{\alpha \\ |I_\alpha|=n}} \prod_{l\in I_\alpha} n_{p_l} \prod_{q\notin I_\alpha} (1-n_{p_q})\label{eq:chiexpand}   \\
  &=\prod_{j} (1- n_{p_j} + e^{i\lambda} n_{p_j}) \label{eq:chidet1} \\
  &= \det\left(1- \hat{n} + e^{i\lambda} \hat{n}\right) \label{eq:chidet2}
\end{align}
after having defined
\begin{align}
\hat{n}&=\sum_{l}n_{p_l}|p_l\rangle\langle p_l| \\
&\to_{R\to\infty} \int \dd k_1 n_{k_1} |k_1\rangle\langle k_1| \ .
\end{align}
Note it is easy to check that \eqref{eq:chidet1} can indeed be expanded into \eqref{eq:chiexpand}. Also, note that \eqref{eq:chidet2} can serve as a continuum definition for $M(\lambda)$. This allows us to remove our regulator and consider again the infinite volume system.

Now let us see how $M(\lambda)$ is related to the entanglement entropy. The key reason behind this relation is that $M(\lambda)$, or rather its logarithm, can be used to generate a spectral representation of any function of $\hat{n}$. Indeed, we have the following formal relations
\begin{align}
 \log(M(\lambda)) &= \Tr\log(1- (1-e^{i\lambda})\hat{n}) \label{eq:logchi} \\
  &=\Tr\left(\log(z- \hat{n}) - \log(z) \right )  \\
  &=\Tr\Big(\int_{0}^z\dd z'\frac{1}{z'- \hat{n}} - \int_1^z \dd z' \frac{1}{z'}  \notag\\
  &\phantom{aaaaaa}-\log(\hat{n})\Big ) \\
  \partial_z   \log(&M(\lambda(z))) = \Tr\left(\frac{1}{z- \hat{n}} -  \frac{1}{z} \right )
\end{align}
with $z=(1-e^{i\lambda})^{-1}$. As a result,  expressions of the type $\Tr(f(\hat{n}))$ for $f$ some function of $\hat{n}$ can be represented as
\begin{align}
  \Tr(f(\hat{n})-f(0)) = \Im\frac{1}{\pi}  \int \dd z f(z) \partial_z   \log\left (M(\lambda(z\pm i\epsilon))\right)
\end{align}
where the $\epsilon$ prescription needs to be chosen appropriately depending on the actual function.

In our case, the probability of creating a pair with momentum $k_1$ is given by the Bogoliubov coefficients $n_{k_1} = \left|\mbeta{k_1}{t^*}\right |^2$. As such we can rewrite the entanglement entropy (\ref{eq:EEntropy}) as
\begin{align}
  S_E &= -\Tr\left ( (1-\hat{n})\log(1-\hat{n}) + \hat{n}\log(\hat{n}) \right) \\
  &= - \Im \frac{1}{\pi} \int_0^1 \dd z \partial_z\log\left (M(\lambda(z- i\epsilon))\right) \label{eq:entropy_spec}\\
    &\phantom{aaaaaaaaaaaaaaaa}\cdot \left ( (1-z)\log(1-z) + z\log(z) \right) \notag \ .
\end{align}
This formula is already interesting as it quantitatively shows that the full knowledge of the distribution of produced particles is enough to infer the entanglement entropy between the particles in the pairs.

This expression can be further expanded to directly relate the entanglement entropy to the moments of the distribution of particles. This is interesting as these are what  would actually be measured in an experiment.  As usual, the moments $C_l$ are obtained from the derivatives of the logarithm of $M(\lambda)$
\begin{align}
  \log  M(\lambda(z)) = \sum_{l=1}^\infty \frac{(i\lambda)^l}{l!} C_l \ . \label{eq:moments}
\end{align}
Integrating by part \eqref{eq:entropy_spec}, inserting \eqref{eq:moments} and recalling that
\begin{align}
  \lambda=-2\left( \frac{\pi}{2} +\frac{i}{2}\log\left(\frac{z}{1-z}\right)\right)
\end{align}
we obtain
\begin{widetext}
  \begin{align}
    S&=\sum_{l=1}^\infty \frac{C_l}{l!}\frac{(-2)^l}{\pi}\Im \int_0^1 \dd z \left( \frac{i\pi}{2} +\frac{1}{2}\log\left(\frac{z}{1-z}\right)\right)^l\log\left(\frac{z}{1-z}\right) \ .
  \end{align}
\end{widetext}
Using the change of variable $u=\frac{1}{2}\log\left(\frac{z}{1-z}\right)$ and the integral \cite{gradshteyn2007}
\begin{align}
  \int_0^\infty\dd u \frac{u^{2l}}{\sinh^2(u)} = \pi^{2l}|B_{2l}|
\end{align}
where $B_{2l}$ are the Bernoulli numbers, we obtain
\begin{align}
  S_E=\sum_{l=1}^\infty \frac{C_{2l}}{(2l)!}(2\pi)^{2l} |B_{2l}| \label{eq:entropyAsExp} \ .
\end{align}
As already mentioned, the same expression was derived for the entanglement entropy produced in quantum shot noise \cite{Klich:2008un} at quantum point contacts. This is not surprising, as the main ingredient in the derivation is the existence of a "full counting statistics" and the fermionic nature of the particles.

Let us further expand this relation. We use the following relation for the Bernoulli numbers \cite{weisstein}
\begin{align}
  B_{2l} = \frac{(-1)^{l-1}2(2l)!}{(2\pi)^{2l}}\zeta(2l)
\end{align}
and rewrite
\begin{align}
  S_E=2\sum_{l=1}^\infty \zeta(2l) C_{2l} \label{eq:entropyCumZeta}
\end{align}
with $\zeta(x)$ the Riemann $\zeta$ function. This form has the advantage of emphasizing the asymptotic nature of the expansion. The $\zeta$ function quickly approaches unity as $l$ grows and the number of contributions to the moments is factorially growing.

Another advantage of this expression is that it can be analytically extended to a potentially convergent form. It can be done as follows. First, write the $\zeta$ function as an absolutely convergent sum for $x>1$,   $\zeta(x)=\sum_{n=1}^\infty\frac{1}{n^x}$ and then formally swap the two summations
\begin{align}
  S_E&=2\sum_{n=1}^\infty\sum_{l=1}^\infty  \frac{C_{2l}}{n^{2l}} \\
  &=2\sum_{n=1}^\infty f_C(n) \label{eq:SESwapped}
\end{align}
where the first equality can be interpreted as a new definition and where we defined  $f_C(x)$ to be the following asymptotic expansion around $x=\infty$
\begin{align}
  f_C(x)=\sum_{l=1}^\infty  \frac{C_{2l}}{x^{2l}} \ .
\end{align}
Because of the generic factorial growth of its coefficients $C_{2l}$, we consider its Borel transform
\begin{align}
  \mathcal{B}f_C (p)=\sum_{l=1}^\infty \frac{C_{2l}}{(2l)!} p^{2l-1}
\end{align}
which generically will have a non-zero radius of convergence. The study of this Borel transform will depend on the specific moments $C_{2l}$. Assuming it does not possess singularities along the positive real axis, it can be used to Borel re-sum $f_C(x)$, see \cite{Costin:1999798} for a review,
\begin{align}
  f_C(x)=\int_0^\infty \dd p e^{-p x} \mathcal{B}f_C (p) \label{eq:BorelLaplace}\ .
\end{align}
In case it does, these singularities can be interpreted as some non-perturbative contributions which can be treated by standard methods, see \cite{Marino:2012zq} for an introduction.

Equations \eqref{eq:SESwapped} and \eqref{eq:BorelLaplace} are more than just formal manipulations; they can be used to provide an efficient way of computing the entanglement entropy from the knowledge of the multiplicity distribution of particles. This can be achieved for instance by using the Pad\'e-Borel methods of \cite{Costin:2019xql,Costin:2020hwg,Costin:2020pcj}.  We exemplify this fact and its improvement over the bare asymptotic expansion \eqref{eq:entropyAsExp} in the next section, where we compute the entanglement entropy in a specific pulsed background from the particle distribution.
We stress again that this approach allows converting a (partial) knowledge of the multiplicity distribution of particles into an (imperfectly reconstructed) entanglement entropy in a practical way.

\section{Sauter pulse:\\ ``entanglement equilibration"}
\label{sec:sauterField}

Let us now apply these results to a specific example. We consider the case of the Sauter pulse
\begin{align}
  A(t)= \Ec\tau\tanh{\totop} \ , \ \ E(t)=\frac{\Ec}{\cosh^2\totop} \ . \label{eq:SauterPulse}
\end{align}
We compute the associated time-dependent Bogoliubov coefficients in appendix~\ref{app:Sauter}.

\begin{figure}[h!]
\vskip0.5cm
  \centering
  \includegraphics{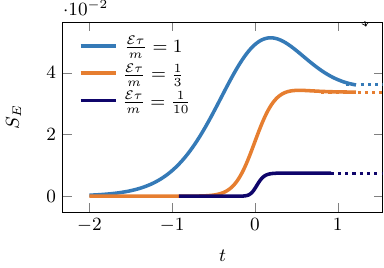}
	\caption{Entanglement entropy as a function of time, shown for different values of the dimensionless ratio $\gamma=\frac{\mathcal{E}\tau}{m}$, where $\mathcal{E}$ is the electric field strength and $\tau$ is the duration of the pulse. We observe that as the pulse is getting shorter, the entropy possesses less distinct features and the equilibration becomes more efficient, see text for discussion.}
	\label{fig:timeDepBogo}
\end{figure}

Let us start by studying the behavior of the entanglement entropy \eqref{eq:EEntropy} associated with the time-dependent Bogoliubov coefficients given by equations~\eqref{eq:alphaSauter}-\eqref{eq:betaSauter}. We show this quantity in figure~\ref{fig:timeDepBogo} for different values of the dimensionless ratio $\gamma=\frac{\mathcal{E}\tau}{m}$. In all cases, we see that the entropy equilibrates to an asymptotic value $S^\infty_E$.
Moreover, as $\gamma$ becomes smaller, the equilibration happens on a shorter and shorter time scale and any distinctive feature gets erased ($\gamma$ is sometimes referred to as the "adiabaticity parameter" \cite{Popov:1971iga}).

\begin{figure*}
	\includegraphics{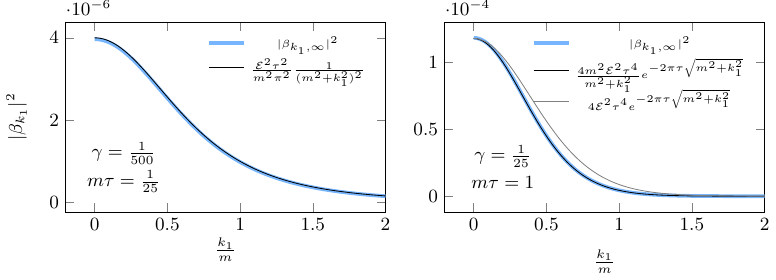}
	\caption{The momentum spectrum of the produced particles. On the l.h.s., we show the the asymptotic spectrum in the regime of small $\gamma$ and small $m\tau$,  while on the r.h.s. we keep $m\tau$ finite. For illustration purposes, in the latter case, we also show the approximated spectrum without the polynomial ``gray factor".}
	\label{fig:betaAs}
\end{figure*}

The asymptotic value for the entropy $S_E^\infty$ can actually be computed from the well known \cite{Narozhnyi:1970uv,Popov:1971iga, Bernard:1977pq,Ambjorn:1983hp}  expressions for the asymptotic coefficients
\begin{widetext}
\begin{align}
	|\malpha{k_1}{\infty}|^2&= \frac{\sinh\left(\pi m\tau(\gamma+\frac{1}{2}(\lambda^++\lambda^-)\right)\sinh\left(\pi m\tau(-\gamma+\frac{1}{2}(\lambda^++\lambda^-)\right) }{\sinh(\pi m\tau\lambda^+) \sinh(\pi m\tau\lambda^-)} \label{eq:asAlphaSauter}\\
	|\mbeta{k_1}{\infty}|^2 &=  \frac{\sinh\left(\pi m \tau(\gamma-\frac{1}{2}(\lambda^+-\lambda^-)\right)\sinh\left(\pi m\tau(\gamma+\frac{1}{2}(\lambda^+-\lambda^-)\right) }{\sinh(\pi m\tau\lambda^+) \sinh(\pi m\tau\lambda^-)}\label{eq:asBetaSauter}
\end{align}
\end{widetext}
with
\begin{align}
	\lambda^{\pm} &= \sqrt{1 + \left(\frac{k_1}{m} \pm \gamma\right)^2}
\end{align}
These expressions can also be obtained as the late time asymptotics of our time-dependent expressions \eqref{eq:alphaSauter}-\eqref{eq:betaSauter}. Given the apparent  fast equilibriation to this asymptotic regime, a legitimate question to ask is how far from thermality this regime is. To find an answer, we expand these asymptotic Bogoliubov coefficients for small $\gamma$. Focusing on $|\mbeta{k_1}{\infty}|^2$, which corresponds to the distribution of created particles we find
\begin{align}\label{dist}
|\mbeta{k_1}{\infty}|^2 &\sim_{\gamma\to 0}  \frac{\gamma^2 m^2\tau^2}{\lambda^2}\frac{1}{\sinh^2(\pi m\tau\lambda)} \\
&\sim \begin{cases}
	\frac{\Ec^2 \tau^2 }{m^2\pi^2}\frac{1}{(m^2+k_1^2)^2} , \ \ \ \ \ \ \ \ \ \ \ \ m\tau <<1 \\
	\frac{4 m^2 \Ec^2\tau^4 }{m^2+k_1^2}e^{-2\pi\tau\sqrt{m^2+k_1^2}}, \ \ m\tau \gtrsim 1
\end{cases}\label{eq:BetaAsExpanded}
\end{align}
with \begin{align}
	\lambda &= \sqrt{1 + \left(\frac{k_1}{m}\right)^2}
\end{align}
and where we explicitly rewrote all the expressions in terms of dimensionful parameters in \eqref{eq:BetaAsExpanded}. We first expand in the limit $\gamma<<1$, $\gamma<< m\tau$ and then consider the two different cases $m\tau <<1, m\tau >> 1$. In the former case, we find the spectrum that decays as $1/(m^2+k_1^2)^2$ while in the latter case the distribution resembles a thermal distribution at temperature $T=\frac{1}{2\pi\tau}$, modulated by a soft ``gray factor". This behavior is illustrated in Fig. \ref{fig:betaAs}.  Note that similar expressions were already discussed in \cite{Popov:1971iga} and the appearance of a "Boltzmann" suppression for short pulses was discussed in \cite{Kharzeev:2005iz}, where references
 to earlier work on phenomenology of hadron production by (chromo)electric fields can also be found.

The momentum distribution (\ref{dist}) is interesting, as it resembles the transverse momentum spectra measured in high energy hadron and heavy ion collisions. Of course, our treatment has been limited to $(1+1)$-dimensional case, but it is known that for short pulses the momentum distribution of produced particles is isotropic \cite{Kharzeev:2005iz}, so the transverse momentum distributions in this case are also given by (\ref{dist}). If we model a high energy collision by an electric pulse, and decompose it in a superposition of Sauter pulses, then the component with the shortest duration $\tau$ would give us a small number (suppressed by $\tau^2$, see (\ref{dist})) of high transverse momentum particles with a power spectrum $\sim 1/(m^2+k_1^2)^2$. The component with $\tau \sim 1/m$ gives rise to the thermal-like component with the exponential transverse momentum spectrum. The effective temperature according to (\ref{dist}) is $T=\frac{1}{2\pi\tau}$, in accord with the semiclassical arguments of \cite{Kharzeev:2005iz}.

\begin{figure}
	\includegraphics{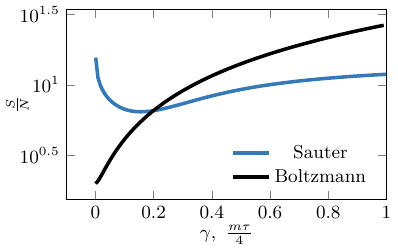}
	\caption{Entanglement entropy for particles produced by Sauter pulse and the thermodynamical entropy. Both quantities are  normalized by the number of particles: for the case of the Sauter pulse, these are the produced particles, and for the thermodynamical entropy the number of particles is computed from the Boltzmann distribution at temperature $T=\frac{1}{2\pi \tau}$. We fix $m=1$ and $\Ec=1$ and plot as a function of $\gamma=m\tau/4$.}
	\label{fig:tauNorm}
\end{figure}

The emergence of an effective thermal behavior is also manifest in  the asymptotic entropy $S_E^\infty$. We show its dependence on $\tau$, at fixed $\Ec$ and $m$ in Fig. \ref{fig:tauNorm}. As we will want to compare it to the thermal case, we normalize by the density of produced particle $N_{prod}=\int \dd k_1 |\mbeta{k_1}{\infty}|^2$. It peaks as $\tau$ goes to zero and asymptote to some finite values for large $\tau$. An intuition behind this behavior is the following. Very short pulses $\tau$ can only probe high frequencies $k\sim \frac{1}{\tau}$, which are very unlikely to be produced. When they are, in our limit, $\Ec<< \frac{m}{\tau}$ and such creation of particle is the result of a large amplitude quantum fluctuation. In such a fluctuation, the momentum conservation imposes a strong entanglement of the produced particles. For longer pulses, the momenta of  the particles are correlated through the electric field, and thus the entanglement entropy per produced particle decreases at early times, see Fig. \ref{fig:tauNorm}.

It is instructive to compare the equivalent quantity obtained from the Boltzmann distribution. In one spatial dimension, it is computed as
\begin{align}
	s &= \frac{p+\epsilon}{T}\\
	\epsilon &= 2\int \frac{\dd p}{2 \pi} \sqrt{p^2+m^2} e^{-\frac{\sqrt{p^2+m^2}}{T}}\\
	p &= 2\int \frac{\dd p}{2 \pi} \frac{p^2}{\sqrt{p^2+m^2}}e^{-\frac{\sqrt{p^2+m^2}}{T}}\\
	n &= 2\int \frac{\dd p}{2 \pi} e^{-\frac{\sqrt{p^2+m^2}}{T}}\\
	\frac{S}{N}&=\frac{s}{n} \label{eq:thermalEntropyNorm}
\end{align}
with $s,\epsilon,p,n$ the entropy, energy, pressure and number density, $T$ the temperature and $S,N$ the total entropy and number of particles. The ratio of the entropy to the number of particles is particularly simple in two limits. When $T\to \infty$, the mass of the particle becomes irrelevant and on dimensional ground we have $s\sim T, n\sim T$, leading to a constant ratio $\frac{S}{N}$. At zero temperature, while pressure goes to zero, we have $\epsilon\sim m n$ so that $\frac{s}{n}\sim \frac{m}{T}$ and it diverges.

We show this expression for $T=\frac{1}{2\pi\tau}$, as suggested by the exponential factor of equation~\eqref{eq:BetaAsExpanded}. For small $\tau$, corresponding to large temperatures, the Boltzmann expression goes to a constant as expected. For finite but small $\tau$, the entanglement entropy per particle for the Sauter pulse is larger than the thermal case. As explained above, this can be understood from the fact that the particles produced by the electric field in this limit result from large amplitude vacuum fluctuations. For $\tau\sim \frac{1}{m}$, the curves cross; for a range of $\tau$, the distribution of created particles is almost thermal. The fact that this happens at $\tau\sim \frac{1}{m}$ is not surprising as this corresponds to pulses that probe modes of order $k\sim m$. Finally, for yet larger $\tau$, the entanglement entropy becomes smaller than in the thermal case; the particles created in pairs are not "maximally random" and thus differ from a thermal state. To sum up, very short pulses are dominated by hard modes and create a distribution that is far are far from thermal. The physics of moderately short pulses is dominated by a Boltzmann suppression factor at an effective temperature of $1/(2\pi \tau)$, leading to an entropy comparable to the one of the Boltzmann distribution. A sharper understanding of the extent to which the Sauter pulse leads to "thermal" states is however still lacking and is left for further work.

To conclude this section, we want to compute again the entanglement entropy, but this time only using the moments $C_{2l}$ of the distribution of created particles. From equation~\eqref{eq:logchi}, we get
\begin{align}
  \log(M(\lambda)) = \int \dd k_1 \log\left(1-(1-e^{i\lambda})|\mbeta_{k_1}{t^*}|^2\right) \ ,
\end{align}
and, correspondingly, we can directly compute the moments from the Bogoliubov coefficients as
\begin{align}
  C_l = \frac{l!}{i^l} \frac{\partial^l}{\partial \lambda^l} \int \dd k \log\left(1-(1-e^{i\lambda})|\beta_k|^2\right) \ .
\end{align}
For completeness, we write down the few first coefficients which contribute to the entropy
\begin{align}
  C_2 &=-4 \int \dd k_1 (|\mbeta{k_1}{t^*}|^4 - |\mbeta{k_1}{t^*}|^2) \\
  C_4 &= 12\int \dd k_1 (-6|\mbeta{k_1}{t^*}|^8+12|\mbeta{k_1}{t^*}|^6\\
	&\phantom{aaaaaaaaaaaaaaaaaaaaaa}-7|\mbeta{k_1}{t^*}|^4 + |\mbeta{k_1}{t^*}|^2) \notag\\
  C_6 &= -720\int \dd k_1 (120|\mbeta{k_1}{t^*}|^{12}-360|\mbeta{k_1}{t^*}|^{10}\\
	&+390|\mbeta{k_1}{t^*}|^8-180|\mbeta{k_1}{t^*}|^6+31|\mbeta{k_1}{t^*}|^4 - |\mbeta{k_1}{t^*}|^2)\notag \ .
\end{align}

\begin{figure*}
  \centering
  \includegraphics{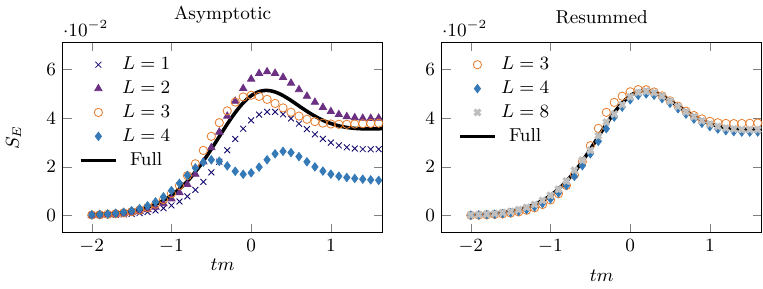}
	\caption{Entanglement entropy as a function of time obtained from the moments of the distribution of created particles. On the l.h.s., we show the entropy obtained from the asymptotic expansion~\eqref{eq:entropyCumZeta} truncated after $L$ terms. On the r.h.s., we show the results obtained with the resummed expression~\eqref{eq:resumFc}.}
	\label{fig:SEMoments}
\end{figure*}

As we are performing some numerical evaluation, we can only evaluate a finite number $L$ of moments. This of course would also be the case if we were to perform an experiment, where only a finite number of cumulants can be measured. Let us start by considering expression \eqref{eq:entropyCumZeta}. We show again the time-dependence of the entropy for some given parameters in the left-hand side of Fig.~\ref{fig:SEMoments}, black line. We also show the results obtained by truncating \eqref{eq:entropyCumZeta} after $L$ terms, for $L=1,2,3,4$. The asymptotic nature of this expansion is very clear. The truncations with $L=1,2,3$ start reproducing the full answer with increased precision  but adding terms beyond $L=3$  makes the expansion break down, as already illustrated by the $L=4$ truncation. In this case, it does not even have the correct qualitative features anymore. Still, it is worth emphasizing that one can get a relatively good idea of what the entanglement entropy is by just using the first moments of the particle distribution.

In order to be able to perform better with more moments and reconstruct the entropy more accurately, we need to resort to the resummation \eqref{eq:SESwapped}. To perform it in practice, we apply the Pad\'e-Borel method of \cite{Costin:2019xql,Costin:2020hwg,Costin:2020pcj}, see also \cite{Florio:2019hzn,Dunne:2021acr} for other concrete examples. It works as follows. First, we construct a truncated Borel transform
\begin{align}
  \mathcal{B}f_C^L (p)=\sum_{l=1}^L\frac{C_{2l}}{(2l)!} p^{2l-1} \ .
\end{align}
With infinitely many terms, the "resummation" is realized by then computing the Laplace transform of the Borel transform. With a finite number of terms, nothing would be achieved if we were to directly apply the Laplace transform. A way to understand this is that the resummation is sensitive to the whole analytic structure of the Borel transform. To perform a resummation with a finite number of terms, we need to interpolate the truncated Borel transform with some functions which can capture its global analytic structure. A practical way of doing this is to use Pad\'e approximants. One can actually show it is the optimal way to do it for meromorphic functions \cite{Costin:2020hwg,Costin:2020pcj}. This leads us to define an interpolated Borel transform
\begin{align}
  \mathcal{PB}f_C^L (p)=\textrm{Pad\'e}_{L,L+1}\big(\mathcal{B}f_C^L (p)\big ) \ ,
\end{align}
where by $\textrm{Pad\'e}_{L,L+1}$ we mean we compute the associated rational  Pad\'e approximant of degree $L,L+1$
\begin{equation}
\textrm{Pad\'e}_{L,L+1}(f)(x)=\frac{\sum_{n=0}^L a_n x^n}{\sum_{m=0}^{L+1} b_m x^m}
\end{equation}
with the coefficients $a_n$ and $b_m$ uniquely determined by the requirements that the Taylor expansions of $f(x)$ and $\textrm{Pad\'e}_{L,L+1}(f)(x)$ needs to match around $x=0$. A priori,  the degree of the approximant can be taken to be $M,N$ with $M+N=2L+1$ but with $M$ and $N$ arbitrary. While sometimes physical arguments can be used to constraint the choice of $M$ and $N$ \cite{Dunne:2021acr}, here we set $M=L$ and $N=L+1$  after some experimentation, realizing that this choice led to particularly stable extrapolations.

Finally, we compute our numerically resummed expression as
\begin{align}
	f_C^L(x)=\int_0^\infty \dd p e^{-p x} \mathcal{PB}f_C^L (p) \ . \label{eq:resumFc}
\end{align}
Note that once the approximation $f_C^L(x)$ is computed, the sum in equation~\eqref{eq:SESwapped} is easy to evaluate.

The results of this resummation are presented on the right-hand side of Fig. \ref{fig:SEMoments}. Curves with only one and two cumulants are not shown, as they do not lead to any useful results. On the other hand, starting from $L=3$, we see an important improvement with respect to the original expansion. And crucially, contrary to the original one, this expansion is convergent; adding higher cumulants only improves the result. This is best illustrated by looking at the results for $L=4$ and contrasting them with the one obtained from the asymptotic expansion, on the left-hand side panel. We strengthen this point by also showing results with $L=8$ to confirm the fact that we can include more and more cumulants to reconstruct the entanglement entropy if we use the resummed expression \eqref{eq:SESwapped}.


\section{Conclusion}
\label{sec:conclu}

In this work, we addressed the question of entanglement between pair-produced particles in a strong electric field. We started by deriving the entanglement entropy between the left- and right-movers (or equivalently between the positive and negative frequency modes) and showed that it matches the statistical Gibbs entropy associated with the created pairs. We then made explicit the link between the multiplicity distribution of created particles and the entanglement entropy by deriving an asymptotic expansion of the entanglement entropy in terms of the cumulants of this multiplicity distribution. We also discussed a re-summation of this expression and argued that it provides a practical method for  reconstructing the entropy from the cumulants.

Finally, we studied a concrete example. We computed the full time-dependent entanglement entropy between left and right movers in a Sauter-type pulsed electric field. We then studied its late-time behavior and pointed out that the resulting spectrum of created particles is strikingly similar to the transverse momentum distribution observed in high energy hadron and heavy ion  collisions, opening further directions to investigate. Indeed, similar ideas have inspired phenomenological model of heavy-ion collision motivated by the phenomenon of pair-production induced by chromoelectric fields, see for instance \cite{Casher:1978wy,Biro:1984cf, Gatoff:1987uf} for early works on these so-called "flux-tubes" models.  We also studied the asymptotic expansion of the entanglement entropy in terms of cumulants. In particular, we implemented the re-summation proposed above and were able to accurately reconstruct the entanglement entropy, greatly improving over the asymptotic expansion.

This work opens up a number of interesting avenues for future research. For instance, the resemblance of the momentum spectra of particles produced by the Sauter pulses to the spectra measured in high energy hadron and heavy ion collisions  and an approximately thermal behavior of multiplicity distributions deserves to be investigated further.  Generalizing our results to interacting theories would also be of great interest.

\subsection*{Acknowledgments}

\appendix

We thank E.~Grossi for interesting discussions. This work was supported by the U.S. Department of Energy, Office of Science, Office of Nuclear Physics, grants Nos. DE-FG88ER40388  and  DE-SC0012704. The work on numerical calculations was supported by the U.S. Department of Energy, Office of Science, National Quantum Information Science Research Centers, Co-design Center for Quantum Advantage (C2QA) under contract number DE-SC0012704.

\section{Plane waves spinors}
\label{app:freeSpinors}

We use this appendix to make our conventions explicit and to write down the explicit form we used for the free spinors. We use the following $\gamma$-matrices
\begin{align}
	\gamma_0 = \begin{pmatrix}
		0& -i\\
		i& 0
\end{pmatrix}, \ \ \  \gamma_1 = &\begin{pmatrix}
    0& i\\
    i& 0
\end{pmatrix}, \ \ \  \gamma_5=\gamma_0\gamma_1 .
\end{align}
We consider the Dirac equation in a constant $A_1$ background
\begin{align}
  \left[i\gamma_0 \dr_0 + \gamma_1(i\dr_1 + A) -m\right]\psi =0  \ .
\end{align}
A generic solution is given by
\begin{align}
  u^\pm_A(t,k_1)= e^{\mp i\omega^\pm t \mp i k_1 x} u_A^\pm(k_1) \ ,
\end{align}
with the following dispersion relation
\begin{align}
  \omega^{\pm\,2} = m^2+(\pm k_1 +A)^2 \ .
\end{align}
We take $u_A^\pm$ to be
\begin{align}
  u^+_A(k_1)&=\frac{1}{\sqrt{2\omega^+}}\begin{pmatrix}
\sqrt{\omega^+ -( k_1 + A_1)} \\
  i\sqrt{\omega^+ + k_1 + A_1}
\end{pmatrix}\label{eq:freeuAp}\\
u^-_A(k_1) &= \frac{1}{\sqrt{2\omega^-}}\begin{pmatrix}
  \sqrt{\omega^- -(k_1 - A_1)}\\
  -i   \sqrt{\omega^- + k_1 - A_1}
\end{pmatrix}\ .
\end{align}
They satisfy the usual identities
\begin{align}
	u^\pm_A(k_1)^\dagger u_A^\pm(k_1)&=1 \label{eq:freeNorm}\\
	u^\pm_A(-k_1)^\dagger u_A^\mp(k_1)&=0 \label{eq:freeOrtho}
\end{align}
and can be used to expand the free Dirac field in a constant background as
\begin{align}
\hat\psi(x)=\int \frac{\dd k_1}{2\pi\sqrt{2}} e^{-ik_1 x}\left(u^+_A(k_1) a_{k_1} + u^-_A(-k_1) b^\dagger_{k_1}\right) \ ,
\end{align}
with $a_{k_1},b_{k_1}$ free creation/annihilation operators satisfying
\begin{align}
  \{a_{k_1}^\dagger,a_{q_1}\}&=\{b_{k_1}^\dagger,b_{q_1}\} = 2\pi \delta(k_1-q_1) \\
  \{a_{k_1},b_{q_1}\} &=   \{a_{k_1},b^\dagger_{q_1}\} = 0 \ .
\end{align}
Note that in the free expansion we grouped particles by energies ($\omega^-(-k_1)=\omega^+(k_1)$) and that the factor which makes the measure Lorentz invariant is absorbed in the spinor normalization.

\phantom{space}

\section{Sauter pulse, explicit formulas}
\label{app:Sauter}
To be able to compute our time-dependent Bogoliubov coefficients in the background~\eqref{eq:SauterPulse}, we need the associated mode functions \eqref{eq:DiracEq}. This problem was originally solved in  \cite{Narozhnyi:1970uv}, and we briefly rederive this solution here. Then, we extract the associated time-dependent Bogoliubov coefficients. We focus on the positive frequency solution as it is the one that is relevant for the Bogoliubov coefficients; the negative frequency solution is obtained in the same way.

Let us write
\begin{align}
	\spinorp{k_1}{t^*}(t)=\begin{pmatrix}
\chi_{1,t^*}^+(t)\\
\chi_{2,t^*}^+(t)
\end{pmatrix} \ ,
\end{align}
leaving the dependence on $k_1$ of $\chi_{1,t^*}^+(t),\chi_{2,t^*}^+(t)$ implicit for conciseness. Then, the Dirac equation \eqref{eq:DiracEq} in the background \eqref{eq:SauterPulse} can be rewritten in components as
\begin{align}
&\ddot{\chi}_{1,t^*}^+(t)+\Bigg[k_1^2+m^2-i\frac{\Ec}{\cosh^2\left(\frac{t}{\tau}\right)}\\
&+\Ec \tau\tanh\left(\frac{t}{\tau}\right)\left( 2k_1+\Ec \tau\tanh\left(\frac{t}{\tau} \right)\right)\Bigg ]\chi_{1,t^*}^+(t) = 0\notag\\
&\chi_{2,t^*}^+(t)=\frac{-\dot{\chi}_{1,t^*}^+(t)+i( k_1+\Ec\tau\tanh\left(\frac{t}{\tau}\right))\chi_{1,t^*}^+(t)}{m} \ \label{eq:diracEqSecondComp} .
\end{align}
The change of variable
\begin{align}
  y=\frac{1}{2}\left(\tanh\left(\frac{t}{\tau}\right)+1\right) \
\end{align}
maps this system into
\begin{widetext}
	\begin{align}
	  -\tau^2\big[m^2+k_1^2+\Ec(4i(-1+y)y\label{eq:sauterDiff}
		&+ 2k_1\tau(-1+2y)+\Ec\tau^2(1-2y)^2)\big]\chi_{1,t^*}^+(y) \\
	  &+4(-1+y)y(1-2y)\chi_{1,t^*}^{+'}(y) - 4(-1+y)^2y^2\chi_{1,t^*}^{+''}(y) =0\notag\\
	  &\chi_{2,t^*}^+(y)=\frac{i\tau( k_1+\Ec(-1+2y)\tau)\chi_{1,t^*}^+(y)+2(-1+y)y\chi_{1,t^*}^{+'}(y)}{m\tau}
	\end{align}
\end{widetext}
where $\chi_{1,2}'$ indicates differentiation with respect to $y$. In this form, this system can be relatively easily solved in terms of the hypergeometric function $\hyp{a}{b}{c}{y}$. A generic solution is given by \cite{ram2020}
	\begin{align}
  &\chi^+_{1,t^*}(y)=(1-y)^{\frac{i\tau}{2}\omega_{out}}\left(y^{-\frac{i\tau}{2}\omega_{in}}c_{1,t^*}\
 \hyp{a}{b}{c}{y}\right . \notag\\
  &\left . +y^{\frac{i\tau}{2}\omega_{in}}c_{2,t^*} \hyp{a-c+1}{b-c+1}{2-c}{y}\right) \ \label{eq:genericSolSauter}
\end{align}
with  the following notation

\newcommand{\oout}{\omega_{out}}
\newcommand{\oin}{\omega_{in}}
\newcommand{\mPiPInf}{\Pi_{out}}
\newcommand{\mPiMInf}{\Pi_{in}}
{
\allowdisplaybreaks
\begin{align}
	\mPiMInf &= k_1 - \Ec \tau \\
	\mPiPInf &= k_1 + \Ec \tau \\
	\oin &= \sqrt{m^2 + \mPiMInf^2} \\
	\oout &= \sqrt{m^2 + \mPiPInf^2} \\
	a &= i\tau\left(\Ec \tau +\frac{1}{2}(\oout-\oin)\right) \\
	b&=1-i\Ec\tau^2+i\frac{1}{2}(\oout\tau-\oin\tau) \\
	c&= 1-i\tau \oin \ .
\end{align}
}
Note that all the dependence on $t^*$, namely the time at which the solutions reduce to the free ones in a constant background, lies in the coefficients $c_{1,t^*},c_{2,t^*}$. They need to be fixed by matching the two solutions.

In our case, we are interested in the solutions which define the vacuum state at asymptotically early times. Using the fact that $\hyp{a}{b}{c}{0}=1$, we \ compute the asymptotic of \eqref{eq:genericSolSauter} as
\begin{align}
	\chi_{1,t^*}^+(t)\sim_{t\to-\infty} \ \ c_{1,t^*}e^{-it\omega_{in}} +c_{2,t^*}e^{it\omega^{in}}  .
\end{align}
To match the free spinor solution \eqref{eq:freeuAp}, we thus need
\begin{align}
c_{1,-\infty}&= \sqrt{\frac{\omega^+ -( k_1 + A_1)}{2\omega^+}}\\
c_{2,-\infty}&=0 \ .
\end{align}
Note that the second component $\chi_{2}$ is easily computed from \eqref{eq:diracEqSecondComp}.

With the full time-dependent mode function at hand, we can now compute the time-dependent Bogoliubov coefficients using relations~\eqref{eq:alphaFromModes}-\eqref{eq:betaFromModes}. After some algebra, we find

\newcommand{\nVar}{y}
\newcommand{\ot}{\omega(t^*)}
\newcommand{\mPi}{\Pi(t^*)}

\begin{widetext}
\begin{align}
	\malpha{k_1}{t^*}&=\frac{m(1-\nVar)^{\frac{i\tau\oout}{2}}\nVar^{-\ \frac{i\tau\oin}{2}}}{2\sqrt{1+\frac{\mPiMInf(\mPiMInf+\oin)}{m^2}}\sqrt{\ot^2+\ot\mPi}}
	\Bigg [ \frac{i(\mPi + \ot)}{2m^2\tau }\frac{ab}{c}\left((1-(1-2y)^2)\hyp{a+1}{b+1}{c+1}{y}\right)\notag\\
	&+\left(1-\frac{(\mPi + \ot)}{m^2}
	\left(
	-\mPi+(y-1)\oin -y\oout
 \right)
	\right)\hyp{a}{b}{c}{y}\Bigg]\label{eq:alphaSauter}\\
	\mbeta{k_1}{t^*}&=-\frac{m(1-\nVar)^{\frac{i\tau\oout}{2}}\nVar^{-\frac{i\tau\oin}{2}}}{2\sqrt{1+\frac{\mPiMInf(\mPiMInf+\oin)}{m^2}}\sqrt{\ot^2-\ot\mPi}}
	\Bigg [ \frac{i(\mPi - \ot)}{2m^2\tau }\frac{ab}{c}\left((1-(1-2y)^2)\hyp{a+1}{b+1}{c+1}{y}\right)\notag\\
	&+\left(1-\frac{(\mPi -\ot)}{m^2}
	\left(
	-\mPi+(y-1)\oin -y\oout
 \right)
	\right)\hyp{a}{b}{c}{y}\Bigg]  \label{eq:betaSauter}
\end{align}
\end{widetext}
where we introduced the extra notation
\allowdisplaybreaks
\begin{align}
	\mPi &= k_1 + A(t^*) \\
	\ot &= \sqrt{m^2 + \mPi^2} \ .
\end{align}

 Some more algebra shows that these expressions indeed asymptote to \eqref{eq:asAlphaSauter}-\eqref{eq:asBetaSauter} for $t\to\infty$.

\bibliographystyle{h-physrev4}

\end{document}